\documentclass[conference]{IEEEtran}

\usepackage{longtable}
\usepackage{cite}
\usepackage{hyperref}
\usepackage{url}
\usepackage{amsmath,amssymb,amsfonts}
\usepackage{algorithmic}
\usepackage{graphicx}
\usepackage{textcomp}
\usepackage{xcolor}
\usepackage{float}
\usepackage{booktabs}
\usepackage[numbers]{natbib}
\usepackage{booktabs}
\usepackage[font=small,textfont=small, figurename=Fig.,tablename=Table]{caption}
\def\BibTeX{{\rm B\kern-.05em{\sc i\kern-.025em b}\kern-.08em
    T\kern-.1667em\lower.7ex\hbox{E}\kern-.125emX}}
\begin{document}

\title{Attacks \& Defenses Against LLM Fingerprinting\\
}

\author{\IEEEauthorblockN{1\textsuperscript{st} Kevin Kurian}
\IEEEauthorblockA{\textit{Oak Ridge National Laboratory} \\
kuriankg@ornl.gov}
\and
\IEEEauthorblockN{2\textsuperscript{nd} Ethan Holland}
\IEEEauthorblockA{\textit{Oak Ridge National Laboratory} \\
hollandeg@ornl.gov}
\and
\IEEEauthorblockN{3\textsuperscript{rd} Sean Oesch}
\IEEEauthorblockA{\textit{Oak Ridge National Laboratory} \\
oeschts@ornl.gov}
}


\maketitle

\begin{abstract}
As large language models are increasingly deployed in sensitive environments, fingerprinting attacks pose significant privacy and security risks. We present a study of LLM fingerprinting from both offensive and defensive perspectives. Our attack methodology uses reinforcement learning to automatically optimize query selection, achieving better fingerprinting accuracy with only 3 queries compared to randomly selecting 3 queries from the same pool. Our defensive approach employs semantic-preserving output filtering through a secondary LLM to obfuscate model identity while maintaining semantic integrity. The defensive method  reduces fingerprinting accuracy across tested models while preserving output quality. These contributions show the potential to improve fingerprinting tools capabilities while providing practical mitigation strategies against fingerprinting attacks.
\end{abstract}


\section{Introduction}


Large language models (LLMs) have become ubiquitous across industries, from customer service chatbots to code generation tools and content creation platforms. As organizations increasingly rely on these models for sensitive applications, the ability to identify which specific model generated a given text—known as LLM fingerprinting—has emerged as a critical security concern. Model fingerprinting threatens user privacy, enables competitor analysis of proprietary systems, and can facilitate targeted attacks against specific model vulnerabilities.

Recent work by Pasquini et al. \cite{pasquini2024llmmap} demonstrated that LLMs exhibit distinctive behavioral patterns that can be exploited for identification through carefully crafted queries. Their LLMmap tool represents the first systematic approach to  fingerprinting for LLM integrated applications, achieving high accuracy in model identification using manually curated queries and establishing fingerprinting as a viable attack vector. However, LLMmap's reliance on hand-crafted queries may not capture optimal discriminative features, and no defensive countermeasures have been developed to protect against such attacks. LLMmap's manual query design approach presents an optimization opportunity—there may exist more effective query combinations that could improve fingerprinting accuracy or efficiency. Simultaneously, the absence of defensive techniques creates an asymmetric security landscape where LLMmap can identify models with high confidence, but defenders lack effective countermeasures to protect model privacy.

We address these gaps through a dual-perspective augmentation of LLMmap. First, we develop a reinforcement learning-based system that views the task as a combinatorial problem to find the best set of queries, automatically discovering optimal query sets for LLMmap, improving upon– manual query selection. Second, we propose a defensive filtering mechanism that uses a secondary LLM to obfuscate the patterns that LLMmap relies on for identification while preserving semantic content.

Our work makes the following contributions: (1) A RL-based query optimization framework that is similar to LLMmap's performance, achieving 93.89\% fingerprinting accuracy but with only 3 queries, representing a 14.2\% improvement over randomly selecting 3 queries from our candidate pool; (2) A semantic-preserving defense mechanism that significantly reduces LLMmap's fingerprinting success while maintaining output quality above 0.94 cosine similarity;  (3) Evaluation framework using cosine similarity metrics to measure the trade-off between LLMmap evasion success and semantic preservation in filtered outputs.

The remainder of this paper is structured as follows. Section II reviews related work in reinforcement learning optimization and LLM security. Section III details our RL-based enhancement to LLMmap and our filtering-based defense against it. Section IV presents experimental results comparing our approaches to LLMmap's baseline performance. Section V discusses limitations and implications, while Section VI outlines future research directions.

\section{Related Work}

The general security implications of large language models have gained  attention as these systems become more prevalent. Recent comprehensive surveys highlight the expanding threat landscape, documenting adversarial embedding attack vectors that exploit model vulnerabilities \cite{schwinn2023adversarialattacksdefenseslarge}. Within this security domain, LLM fingerprinting has emerged as a critical concern, with several approaches developed for model identification and ownership verification. ImF introduces implicit fingerprinting techniques that embed ownership information through Chain-of-Thought prompting strategies, demonstrating how specific prompting patterns can reveal model identity \cite{jiaxuan2025imfimplicitfingerprintlarge}. Similar to LLMmap, \cite{yang2024fingerprintlargelanguagemodels} model LLM outputs as unique vector spaces, enabling ownership authentication through statistical analysis of response patterns. 

The intersection of reinforcement learning and security tasks has shown promising results, with studies demonstrating RL's effectiveness in adversarial settings. Recent work on cryptographic challenges shows that RL can  improve LLM performance on security-related tasks, supporting the viability of RL-based approaches for complex security applications \cite{muzsai2025improvingllmagentsreinforcement}. The deployment of RL agents in adversarial settings raises concerns about reward hacking, where agents exploit reward signals in unintended ways \cite{Chan_2023}. This informed our reward function design to ensure genuine optimization rather than just pattern exploitation.

Reinforcement learning has proven effective for combinatorial optimization, with Mazyavkina et al. establishing the theoretical foundation for using RL to automate heuristic search in NP-hard problems, while CO-Bench demonstrates that structured RL approaches consistently outperform simpler baselines on complex selection tasks \cite{mazyavkina2021reinforcement}, \cite{sun2025co}. Recent advances have further demonstrated the effectiveness of combining RL agents with natural language embedding spaces, showing that language models can effectively process domain embeddings as input \cite{tennenholtz2024demystifyingembeddingspacesusing} and that language-guided exploration proves effective for agent decision-making in semantic spaces \cite{golchha2024languageguidedexplorationrl}, while approaches for variable action spaces demonstrate how embedding representations can handle discrete selection tasks \cite{sinii2024incontextreinforcementlearningvariable}. Such studies support the idea of incorporating query embeddings directly into the RL agent's state representation for fingerprinting optimization.

On the defensive side, utilizing and modifying vectorized embeddings to detect and mitigate adversarial attacks on models has proven effective \cite{goyal2023surveyadversarialdefencesrobustness}.  
Work on paraphrase attacks against language models proves especially relevant, as it relates directly to the obfuscation and rewording of model-generated text \cite{zhou2024paraphraseidentificationdeeplearning}.

\section{Methodology}
\subsection{Attack: Reinforcement Learning-Based Query Optimization}
\subsubsection{Initial Query Pool}
The foundation of our attack method relies on a diverse and comprehensive query pool designed to elicit discriminative responses from different LLMs. We generate queries across multiple categories that target distinct aspects of model behavior. The process starts with a meta-model approach where we employ a  high-capacity LLM (LLaMa 3.3 70B) to generate category-specific query sets. We define five distinct query categories, inspired by and leveraging \cite{pasquini2024llmmap} for prompting LLaMa outlined below. The entire query pool of 50 queries can be seen in Table \ref{tab:query_pool}
\begin{enumerate}
    \item Meta-Information: Probe for model-specific metadata and self-knowledge.
    \item Alignment Probing: Target model-specific alighnment approaches and safety behaviors
    \item Technical Capability: Assess model-specific technical competencies
    \item Execution Triggers: Queries that bypass vulnerabilities and prompt injection resistance:

\end{enumerate}
This pool will serve as the candidate set of queries from which an optimal set will be determined via a trained agent. 

\subsubsection{Dataset}

Our dataset construction follows LLMmap's methodology while extending it for reinforcement learning-based optimization. We construct a  fingerprinting dataset that captures model behavioral variations across different prompting configurations. We evaluate our approach on a diverse set of open-source language models accessible through Ollama, including variants from major model families: Mistral (7B-Instruct v0.1, v0.2, v0.3), Gemma (1.1-2B-it, 1.1-7B-it), Qwen2 (1.5B-instruct), Aya-23 (8B), SmolLM2 (1.7B), and SOLAR (10.7B-Instruct-v1.0). 

To maintain compatibility with Ollama’s API interface, our configuration space focuses exclusively on hyperparameter variations rather than the full configuration universe used in LLMmap \cite{pasquini2024llmmap}. Our configuration space includes temperature values from 0.0 to 1.0 and frequency penalties from 1.0 to 1.5 and then sample 75 random hyperparameter configurations to prompt the queries. Each trace entry contains the query text, hyperparameter configurations, model responses. The complete dataset comprises responses from our 50-query pool across all model-hyperparameter combinations, resulting in roughly 33,000 of query-response pairs.  While LLMmap uses a fixed set of 8 manually curated queries across a full configuration space, our dataset enables dynamic query selection from a larger pool of 50  generated queries within the hyperparameter-constrained space, allowing for more targeted  fingerprinting strategies that adapt to specific model characteristics.
\subsubsection{Problem Formulation \& Reinforcement Learning Framework}
We formalize the LLM fingerprinting problem as a sequential decision-making task. Given a set of candidate queries $Q = \{q_1, q_2, \ldots, q_n\}$ generated as outlined earlier, and a target model $M$ from model space $\mathcal{M} = \{M_1, M_2, \ldots, M_k\}$, the objective is to select an optimal query subset $Q^* \subseteq Q$ that maximizes classification accuracy:
\begin{equation}
Q^* = \arg\max_{Q' \subseteq Q} P(M = M_i \mid R(Q', M))
\end{equation}

where $R(Q', M)$ represents the response set obtained by querying model $M$ with query subset $Q'$.

We model the query optimization problem as a Markov Decision process with the following components:

\textbf{State Space:} The state $s_t \in \mathcal{S}$ at timestep $t$ is represented as a high-dimensional vector combining:
\begin{equation}
s_t = [|Q_t|, \mathbf{E}_{Q_t}, \mathbf{H}_t]
\end{equation}

where:
\begin{itemize}
    \item $|Q_t|$ is the current query count (scalar)
    \item $\mathbf{E}_{Q_t} \in \mathbb{R}^{20 \times 1024}$ contains embeddings of selected queries (flattened to 20,480 dimensions)
    \item $\mathbf{H}_t \in \mathbb{R}^{36}$ represents the action history (12 timesteps $\times$ 3 components)
\end{itemize}

The 12-timestep episode limit prevents unnecessarily high query sets while allowing time for the agent to still build query sets. Each timestep enables only one query addition but classifier performance evaluation occurs only at episode termination, creating a sparse reward signal that requires the agent to plan query selection strategies without intermediate feedback. The total observation space dimensionality is 20,517, incorporating semantic query representations directly from LLMmap's embedding space\cite{pasquini2024llmmap}.

\textbf{Action Space:} We employ a discrete action space with $2n$ possible actions, where $n = |Q|$ is the size of the query pool:
\begin{equation}
\mathcal{A} = \{\text{ADD}_i : i \in [0, n-1]\} \cup \{\text{NO\_ACTION}_i : i \in [n, 2n-1]\}
\end{equation}

This formulation allows the agent to either select a specific query from the pool or terminate the episode.

\textbf{Action Masking:} To ensure valid actions, we implement action masking:
\begin{equation}
\text{mask}[i] = \begin{cases}
\text{True} & \text{if } q_i \notin Q_t \text{ and } |Q_t| < \text{max\_queries} \\
\text{False} & \text{otherwise}
\end{cases}
\end{equation}

for ADD actions, while NO\_ACTION is always valid.

\textbf{Reward Function}: The reward function balances accuracy maximization with query efficiency, designed to encourage the agent to achieve high fingerprinting accuracy with minimal queries. Our piecewise reward function is:

\begin{equation}
R(s_t, a_t) = \begin{cases}
-5.0 & \text{if } |Q_t| = 0 \\
R_{\text{optimal}}(s_t) & \text{if } |Q_t| \leq 8 \\
R_{\text{transition}}(s_t) & \text{if } 8 < |Q_t| \leq 12 \\
R_{\text{penalty}}(s_t) & \text{if } |Q_t| > 12
\end{cases}
\end{equation}

For the optimal zone ($|Q_t| \leq 8$), motivated by LLMmap's target of 8 queries \cite{pasquini2024llmmap}:

\begin{align}
R_{\text{optimal}}(s_t) &= \alpha \cdot A_t^{2.5} \cdot 5.0 - \beta \cdot |Q_t| \cdot p_t + \gamma \cdot E_t - \delta \cdot P_{\text{acc}}
\end{align}

where:
\begin{itemize}
    \item $A_t$ is the classification accuracy at timestep $t$
    \item $p_t = \min(1.0, \frac{A_t - 0.945}{0.02})$ is a smooth penalty factor
    \item $E_t$ represents efficiency bonuses based on accuracy thresholds
    \item $P_{\text{acc}}$ applies accuracy penalties for suboptimal performance
\end{itemize}

\begin{figure}[htbp]
\centerline{\includegraphics[width=0.5\textwidth]{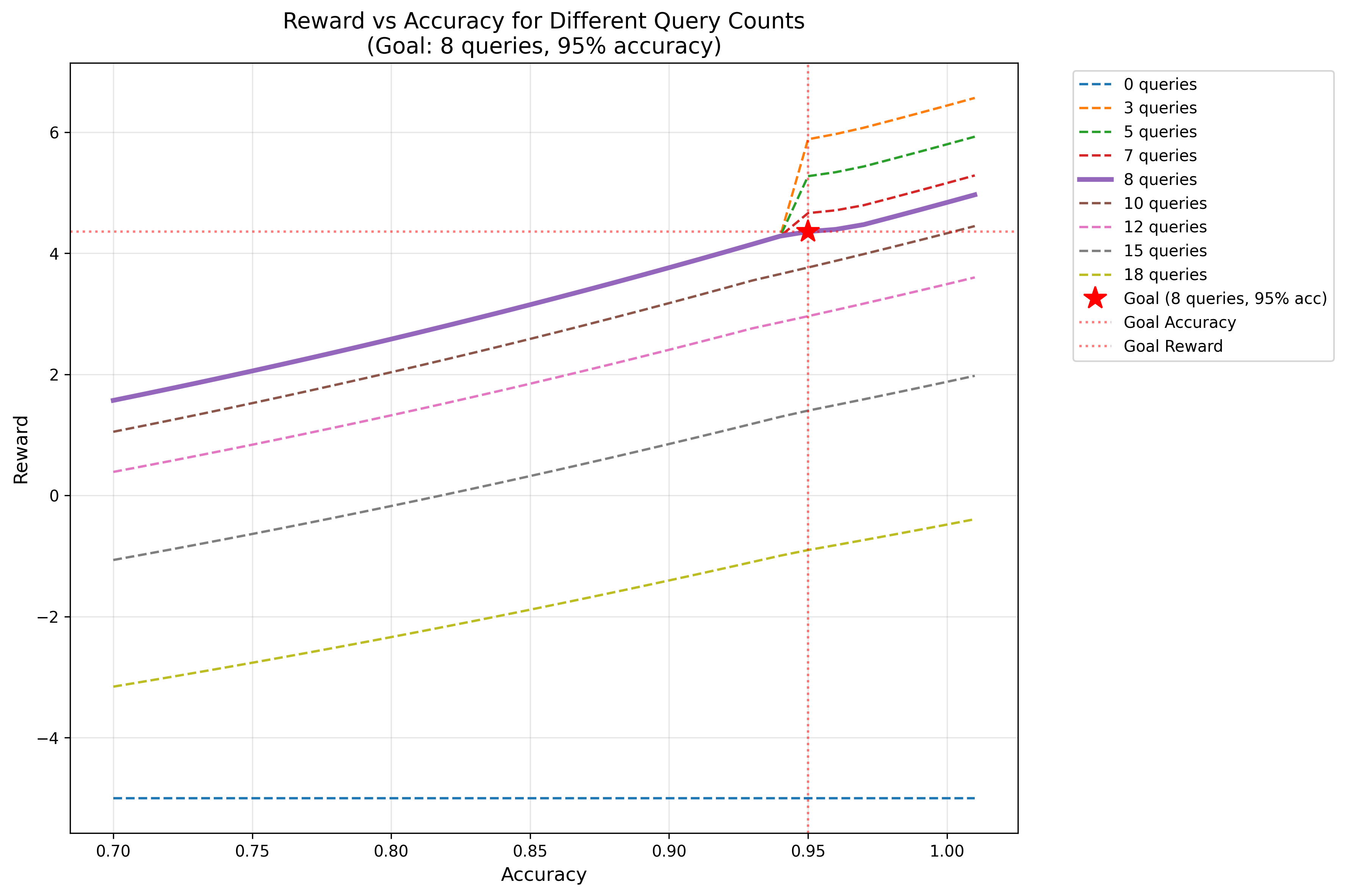}}
\caption{{RL agent performance showing reward versus accuracy trade-off across different query counts. The goal of 95\% accuracy with 8 queries (red star) is achieved with a reward of approximately 4.36, demonstrating optimal query selection efficiency.}}
\label{fig:rewardSpace}
\end{figure}

The transition and penalty zones employ exponential penalties to discourage excessive query usage:

\begin{align}
R_{\text{transition}}(s_t) &= \alpha_w \cdot A_t^{2.5} \cdot 5.0 - (|Q_t| - 8)^{1.3} \cdot 0.15 - P_{\text{acc}} \\
R_{\text{penalty}}(s_t) &= \alpha_w \cdot A_t^{2.5} \cdot 5.0 - (|Q_t| - 15)^{1.4} \cdot 0.25 - P_{\text{acc}}
\end{align}

where $\alpha_w = \max(0.5, 1.0 - (|Q_t| - 8) \cdot 0.03)$ reduces the accuracy weight as query count increases beyond the optimal range. Figure \ref{fig:rewardSpace} shows what the reward function looks like when mapped out across both components. 

Our environment interfaces directly with LLMmap's inference pipeline. The optimizer instantiates LLMmap's transformer architecture with the same hyperparameters outlined in \cite{pasquini2024llmmap}. At episode termination, the agent's selected query set $Q_t$ triggers fresh classifier training. 

This process resets model weights, trains on the selected queries for up to 30 epochs with early stopping, and returns test accuracy for reward computation.  To accelerate training,  query set caching is used to prevent redundant retraining for identical query combinations. We employ Masked PPO with the following hyperparameters and architecture:
\begin{itemize}
    \item Network architecture: [256, 128, 64] 
    \item Learning rate: $7 \times 10^{-4}$
    \item Discount factor: $\gamma = 0.99$
    \item Steps per update: $n_{\text{steps}} = 3072$
    \item Clip range: $\epsilon_{\text{clip}} = 0.2$
    \item Training epochs per update: $n_{\text{epochs}} = 5$
\end{itemize}

The masked action distribution ensures only valid actions receive non-zero probability:
\begin{equation}
\pi_\theta(a_t|s_t) = \frac{\exp(f_\theta(s_t)[a_t]) \cdot \text{mask}[a_t]}{\sum_{a' \in \mathcal{A}} \exp(f_\theta(s_t)[a']) \cdot \text{mask}[a']}
\end{equation}

This architecture enables the agent to learn complex query selection patterns while respecting the combinatorial constraints of the fingerprinting task,  discovering query sets that are capable of LLMmap's baseline performance through  exploration of the semantic query space.

\subsection{Defense: Model Filtering}
To defend against fingerprinting methods we have devised a defensive method that aims to mislead LLMmap while retaining the semantic integrity of the LLM's response. This is implemented through a separate LLM serving as a "filter" model that is instructed to filter out and reword sensitive information. To accurately gauge the performance of our defensive methods a baseline was gathered using LLMmap. We recorded models that LLMmap could consistently identify successfully using its closed-set inference model, running LLMmap on each model 20 times (Table \ref{tab:closed_set_eval}). 

\begin{table}[h]
\begin{center}
\caption{Model Evaluation Scores (Closed-Set). LLMmap achieves near-perfect identification across tested models, establishing a strong baseline for evaluating defensive countermeasures.}
\label{tab:closed_set_eval}
\begin{tabular}{||l|c||}
    \hline
    \textbf{Model} & \textbf{Score} \\
    \hline
    Aya-23-8B & 20/20 \\
    Qwen2-1.5B-instruct & 20/20 \\
    Gemma-1.1-2b-it & 20/20 \\
    Gemma-1.1-7b-it & 20/20 \\
    smollm2:1.7b & 18/20 \\
    upstage/SOLAR-10.7B-Instruct-v1.0 & 20/20 \\
    mistralai/Mistral-7B-Instruct-v0.1 & 20/20 \\
    mistralai/Mistral-7B-Instruct-v0.2 & 20/20 \\
    mistralai/Mistral-7B-Instruct-v0.3 & 20/20 \\
    \hline
\end{tabular}
\end{center}
\end{table}


This leads us to the evaluation of our defensive methods against our collected baseline. However, evaluating the performance of our defenses based solely on LLMmap's accuracy can be quite misleading. For example, a filter model that produces significantly different output from the original model will likely cause LLMmap to fail at identifying the model, suggesting a successful defense. This success is not meaningful however as the semantic integrity of the original output has been completely lost. To address this issue an objective metric is needed to evaluate the filter model's output in relation to the original output, the most appropriate of which being the use of cosine similarity score. Cosine similarity is a measure of semantic difference between strings. The closer to zero the cosine similarity, the more semantically different the two strings are, the closer to one, the more semantically similar two strings are. Using the cosine similarity between the two strings we are able to evaluate the most appropriate filter model prompt by comparing LLMmap's fingerprinting rate to the average cosine similarity score between models. 

The model that served as the filter model was randomized between calls to LLMmap. The models were randomly selected from the pool of baseline models, but the filter model will never be the same as the original model, whose output it is modifying. 

The prompts (Tables \ref{tab:filter_prompts_1}, \ref{tab:filter_prompts_2}) were evaluated (Table \ref{tab:filter_model_prompt_eval}) and and score of 0 to 1 was generated (Eq. \ref{eq:score}), in which the average cosine similarity and average fingerprinting accuracy are taken from trials in which each filter model prompt was tested 20 times with each model in our baseline pool. A score closer to one indicates both success in preventing fingerprinting and retaining semantic integrity of the output. 

\begin{equation}
\label{eq:score}
\text{Score} = 0.5 \times S_{\text{cosine}} + 0.5 \times (1 - A_{\text{fingerprint}})
\end{equation}

\begin{table}
\begin{center}
\caption{Filter Model Prompt Evaluation. Prompt 6 achieves the highest performance score of 0.8562 with 24.4\% correct fingerprinting and 95.6\% cosine similarity. Fingerprinting rates vary substantially across prompts from 24.4\% to 47.4\%, indicating prompt sensitivity in defensive performance.}
\label{tab:filter_model_prompt_eval}
\resizebox{\columnwidth}{35pt}{%
\begin{tabular}{|c|c|c|c|c|}
\hline
& Corr. Fingerprint \%\ & Filter Model ID'd & Avg. Cos. Sim. & Prompt Eval. Score \\
\hline
Prompt 1 & .311 & .172 & .968 & .8283 \\
Prompt 2 & .474 & .152 & .968 & .7467 \\
Prompt 3 & .366 & .188 & .944 & .7888\\
Prompt 4 & .366 & .172 & .955 & .7942\\
Prompt 5 & .305 & .2 & .96 & .8275 \\
Prompt 6 & .244 & .172 & .956 & .8562\\
Prompt 7 & .311 & .177 & .958 & .8238 \\

\hline
\end{tabular}
}
\end{center}
\end{table}

\section{Results}
\subsection{RL Agent Training Results}
\begin{figure}[htbp]
\centerline{\includegraphics[width=0.5\textwidth]{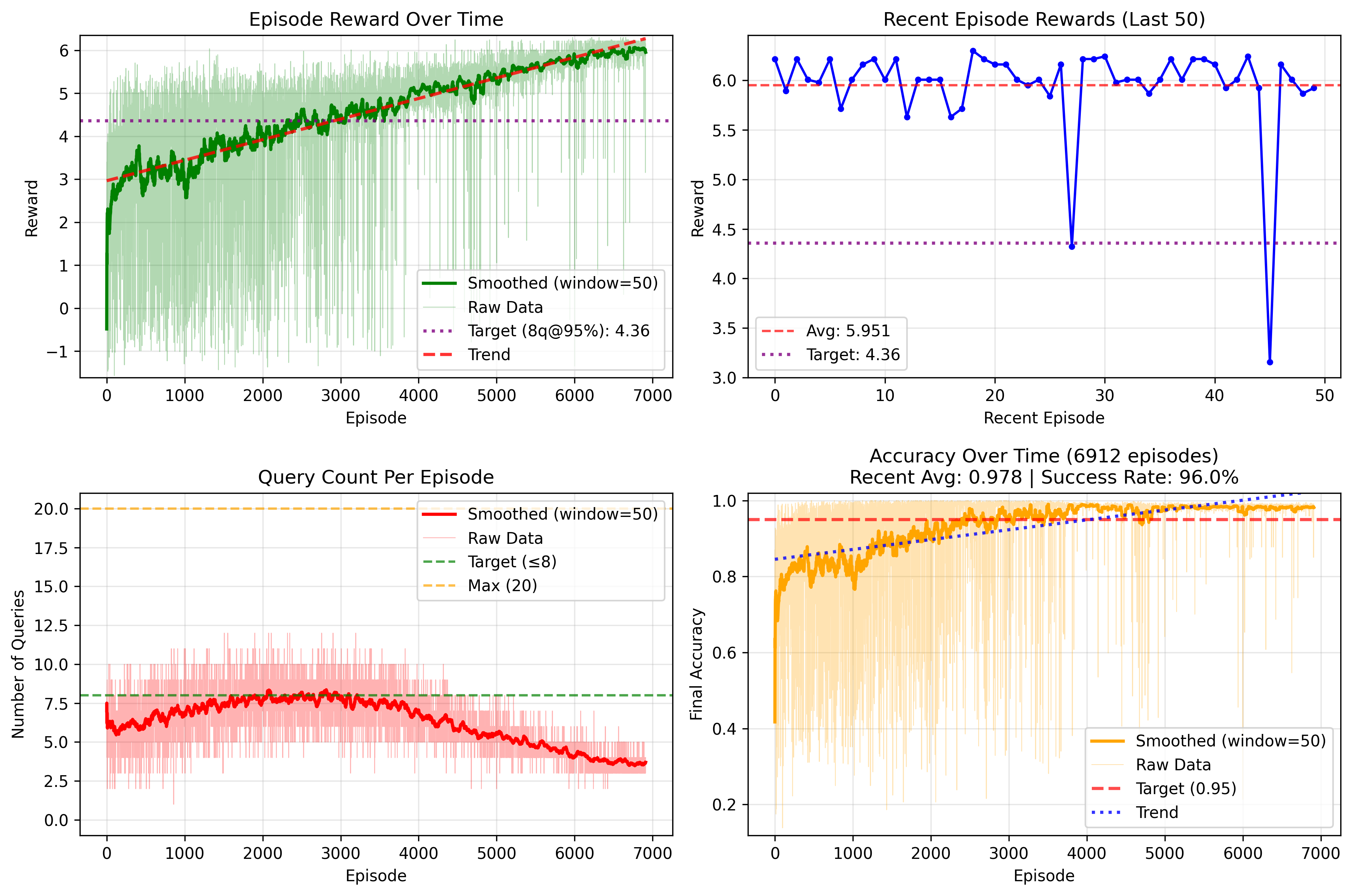}}
\caption{RL training  demonstrating agent convergence after $\sim$7000 episodes. Final performance achieves 97.8\% accuracy using only 3 to 4 queries on average, surpassing both the accuracy target (95\%) and efficiency goal ($\leq$8 queries).}
\label{fig:rl_training}
\end{figure}

During training, the agent achieved convergence after 80,000 timesteps achieving close to 98\% accuracy with only 3 queries as shown in Figure \ref{fig:rl_training}. The agent was able to achieve high performance with much fewer than 8 queries, showing practical implications in that lower interactions with target models minimizes detection probability. During training, we deliberately constrained the train/test split to 35/65 ratio instead of conventional ratios. This unconventional split was motivated by two factors: first, to encourage better generalization given our limited configuration space, second, to force the agent to identify queries that remain discriminative across hyperparameter settings rather than memorizing training-specific patterns to make training harder but still achievable, ensuring the agent learns robust query selection strategies under challenging conditions. 

\subsection{RL Agent Evaluation \& Comparison}
To validate the effectiveness of our RL-optimized query sets, we conduct a  evaluation comparing the agent-selected queries against randomly selected baseline queries of equivalent size. Our evaluation process involves two primary phases: first, we extract the optimal query set discovered by the trained RL agent during the convergence phase, which consistently identified a 3-query combination across all episodes. Second, we compare the optimized query set to a random baseline of equal size to isolate the effect of intelligent query selection.

For classifier training during evaluation, we employ LLMmap's transformer architecture with identical hyperparameters as used during RL training. However, we extend the training duration to 50 epochs for further precision.The evaluation protocol tests each query set against all nine target models with 20 independent fingerprinting attempts per model.

Table~\ref{tab:rl_evaluation} presents the evaluation results comparing RL-optimized queries against randomly selected baselines. The RL-optimized query set achieves an overall accuracy of 93.89\% compared to the random baseline's 82.22\%, representing a relative 14.2\% improvement in fingerprinting. This improvement is notable given that both query sets contain identical numbers of queries (3), isolating the impact of intelligent selection from query quantity effects. Its also important to note that ~80\% accuracy from the random baseline suggests that the query pool itself was fairly strong.

\begin{table}[h]
\caption{RL-Optimized vs. Random Query Set Performance Evaluation. Smollm2 shows the largest performance gap (13/20 vs 5/20), while some models like Gemma-2B and Aya-8B achieve perfect scores regardless of query selection method.}
\label{tab:rl_evaluation}
\begin{center}

\begin{tabular}{||l|c|c||}
    \hline
    \textbf{Model} & \textbf{RL-Optimized} & \textbf{Random Baseline} \\
    \hline
    mistral:v0.1 & 20/20 & 15/20 \\
    mistral:v0.2 & 20/20 & 18/20 \\
    mistral:v0.3 & 20/20 & 17/20 \\
    solar:10.7b & 20/20 & 19/20 \\
    smollm2:1.7b & 13/20 & 5/20 \\
    gemma:7b-instruct & 17/20 & 18/20 \\
    gemma:2b-instruct & 20/20 & 20/20 \\
    qwen2:1.5b-instruct & 19/20 & 16/20 \\
    aya:8b & 20/20 & 20/20 \\
    \hline
    \textbf{AVERAGE} & \textbf{169/180} & \textbf{148/180} \\
    \hline
\end{tabular}
\end{center}
\end{table}

Overall, results show reinforcement learning-based query optimization is capable  of improving LLM fingerprinting effectiveness beyond traditional manual approaches and is a step in the right direction toward automated query generation. 
\subsection{Filter Model Results }

\begin{figure}[htbp]
    \centering
    \centerline{\includegraphics[width=0.5\textwidth]{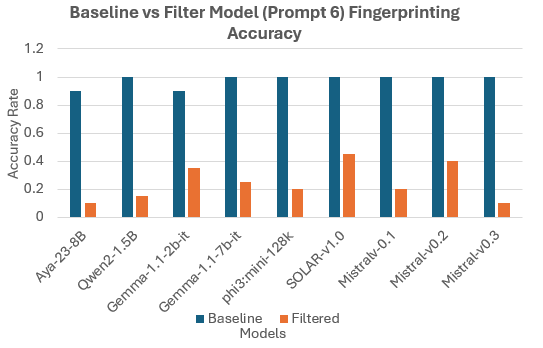}}
    \caption{The filter model reduces fingerprinting accuracy from near-perfect baseline performance (90-100\%) to significantly lower rates (5-45\%) across all models. Aya:8b and mistral:v0.3 show highest defensive effectiveness with filtered accuracy dropping to approximately 20\% and 45\% respectively, while Solar proves most resilient to filtering at 45\% accuracy.}

    \label{fig:defense_filters}
    
\end{figure}

When tested with an appropriate prompt our defensive implementation yields promising results, significantly lowering the accuracy rate of LLMmap (Fig. \ref{fig:defense_filters}). The defense reduces LLMmap's baseline accuracy on the selected pool of models, which ranges from 90-100\%, to anywhere from 5-45\% depending on the model.

Our defensive implementation varies in effectiveness due to several factors, those being the prompt provided to the filter model and the initial model that is being obfuscated, ranging from 5\% accuracy at best to 45\% accuracy at worst. Additionally, the chosen filter model very likely significantly impacts the effectiveness of the defense as well, but there has not been comprehensive testing to prove this claim. 

\section{Limitations}

The RL approach, while having promising results, does have some constraints that limit its effectiveness for a fully automated fingerprinting process. The agent operates in a constrained configuration space and is only evaluated on 9 target models, which raises concerns about generalizability as the ecosystem of LLMs expands. While the current evaluation showed effectiveness within the scope of its training, it would be important to see how the trained agent would perform when confronted with new query pools that are outside of its distribution. The reward function design may be over-engineered when only focusing on two objectives, which may not hold up if the task varies and may lead to training instability. Rather than generating novel queries from an underlying token space, the system is limited by its reliance on a fixed-size, predefined query pool, preventing the discovery of potentially more discriminative query formulations that could emerge from direct generation, however such an approach would add a significant amount of computational cost. Despite incorporating embeddings in the observation space, the agent lacks a very deep understanding of what makes queries discriminative or consistent, moreso relying on statistical patterns instead of fundamental model differences. The agent also requires retraining when introducing new models or query types, making it hard to apply in real-time environments. 

These RL optimization limitations directly compound the challenges faced by downstream defense mechanisms, as suboptimal query selection can make fingerprinting attacks more detectable and predictable. The filter model inherits these upstream constraints and introduces additional drawbacks of its own. The most significant of which being the core structure of the initial output is commonly modified, semantic integrity is retained but the exact wording of the original output is often not, meaning that the exact output is lost in the pursuit of evading identification. While this does not hamper user experience it may not serve specific goals.

Another potential defensive limitation is revealed upon repeated use of fingerprinting tools as a pattern may start to emerge, for example the original model or output model may be identified more often than any other model even if it's only correctly identified 30\% of the time. For one off tests this should not present any issue, however if many tests are being performed a pattern may emerge and an assumption about the true model responsible for output could potentially be made. 
\section{Related Methods and Future Work}


Several other avenues were explored related to defending against fingerprinting attacks but failed to yield significant results. One of which being the inclusion of a third model to serve as a judge of the text generated by the filter model. It would grade the filter model's output based on attributes and instructions given through its prompt. The purpose of the judge would be to evaluate whether or not the filter model's output is semantically and stylistically consistent with the original models output. While this approach showed promise it was impractical for several reasons. The first being that it was quite inconsistent in its scoring of the provided prompts, however, with more effort it would be possible to fine tune the judge through prompt engineering to yield much more consistent and appropriate scores. The version with the judge also took a significant amount more time to run, considering each prompt was run through three models. Though the reason the judge proved impractical was due to the fact that it offered little that the cosine similarity score didn't already. 

Future work should address the limitations identified in both the RL optimization and defense mechanisms. The RL agent could be improved to handle open-set fingerprinting scenarios adding unseen models without further retraining, while expanding the configuration space to include system prompts and different framework architectures beyond the current constrained setup. The reward function should incorporate intermodel discrepancy and intramodel consistency \cite{pasquini2024llmmap} as additional signals to improve query discrimination. Another direction involves the agent moving away from fixed pools and instead through a generative query space in a computationally feasible manner. 

For filter model defense, to further ensure appropriate output, a threshold could be enforced in the case that the output produced a low cosine similarity score, in this case it could be regenerated until its output produces text with a high enough cosine similarity score. However, this may prove to be costly if the output model is repeatedly prompted. 

The limiting factor of the current implementation of the filter model lies in its prompt. There are likely many prompts that would better optimize the ratio of cosine similarity to fingerprinting accuracy. The filter model is  what is responsible for the performance of the model and should thus be optimized. 

Another important facet of this implementation to consider is the filter model used. In the current implementation it is randomized between calls to LLMmap, however, more testing may reveal an optimal filter model. Another promising approach is the use of a smaller language model as the filter model. The current implementation of the filter model may be overkill in many ways given that we essentially only need to reword the input text. A smaller, more specialized, filter model may be much more appropriate to optimize our implementation's use of space and time.

\bibliographystyle{ACM-Reference-Format}
\bibliography{references}
\section{Appendix}
\clearpage
\renewcommand{\tablename}{Table}
\renewcommand{\figurename}{Figure}
\renewcommand{\thetable}{A\arabic{table}}
\renewcommand{\thefigure}{A\arabic{figure}}
\setcounter{table}{0}
\setcounter{figure}{0}

\begin{table*}[t]
\begin{center}
\caption{Filter Prompts 1-4. These prompts represent the first iteration of defensive filtering instructions, ranging from comprehensive rewording guidelines (Prompt 1) to minimal intervention approaches (Prompt 4). Prompt 1 provides the most detailed instructions for preserving semantic similarity while obscuring fingerprinting patterns, while Prompt 4 offers the most concise filtering directive.}
\label{tab:filter_prompts_1}
\renewcommand{\arraystretch}{1.2} 
\small  
\begin{tabular}{|p{0.15\textwidth}|p{0.8\textwidth}|}  
\hline
\textbf{Prompt 1} &
\begin{minipage}[t]{\linewidth}
Given the following prompt: "\{prompt\_text\}"\\

            Your task is to reword the reference answer below if necessary. It is necessary if there is sensitive information in the response that could be used to fingerprint the model used to generate it. You must subtly tweak the answer so that the user does not know while preventing fingerprinting.\\

            Remove or replace any identifying details (e.g., model names, API behaviors, unique stylistic features).\\
            Introduce plausible, misleading alternatives in place of removed information.\\
            Retain the overall meaning, logic, and structure.\\

            Do not change the length of the original response significantly. There must not be a significant amount of new information, hold true to the original.\\
            The response should not be stylistically departed from the original, it should feel as though, if changed at all, it was simply to lightly reword rather than restylize the statement.\\
            Ensure the revised response appears as if it was the original, without any mention of editing, modification, or your role.\\

            It is absolutely forbidden to mention this editing process in any way. Do not say "sure, I can do that!" or anything similar. Only output the modified version of the original response below. We absolutely do not want the user to know about your role. \\
            Aim to preserve semantic similarity (including vector embedding proximity), while making the output look different enough to obscure the original's fingerprint.\\

            Do not change the structure of the original response, it should be very similar in length, style and semantics. Ideally every word would not be a syllable of itself, only some would be reworded, not everything. \\

            Original Response: "\{first\_output\}"\\

            Now produce the revised response: \\
\end{minipage} \\
\hline
\textbf{Prompt 2} & 
\begin{minipage}[t]{\linewidth}
Given the prompt: "\{prompt\_text\}" \\

            And the model's original response: "\{first\_output\}" \\

            Your task is to revise the response only where necessary to remove or obscure any identifiable patterns or content that could be traced back to the model.\\

            Preserve the original response's intent, meaning, and flow.\\
            Only alter or remove details that might be fingerprintable.\\

            Do not add new content or significantly alter the style or length.\\
            Avoid any mention of modification, rewording, or your role.\\

            Output the revised response only. No explanations.\\
\end{minipage} \\
\hline
\textbf{Prompt 3} & 
\begin{minipage}[t]{\linewidth}
You are tasked with anonymizing an LLM-generated response.\\

            Prompt: "\{prompt\_text\}"\\
            Original Answer: "\{first\_output\}"\\

            Your objective is to remove or disguise any details that could reveal which model wrote it, without changing the actual meaning. Maintain the sentence structure and natural tone.\\

            You may lightly reword, but do not rephrase entirely. Avoid overt changes in tone or vocabulary. Do not acknowledge that this was rewritten.\\

            Revised Response:\\
\end{minipage} \\
\hline
\textbf{Prompt 4} &
\begin{minipage}[t]{\linewidth}
Modify the following response to avoid including any model-identifying details while preserving its full semantic content. Keep it stylistically similar.\\

            Prompt: "\{prompt\_text\}"\\
            Original Output: "\{first\_output\}"\\

            Output the revised version only, with no extra commentary.\\
\end{minipage} \\
\hline
\end{tabular}
\end{center}
\end{table*}

\begin{table*}[t]
\begin{center}
\caption{Filter Prompts 5-7. These prompts explore alternative defensive strategies, with Prompt 6 achieving optimal performance (0.8562 score) through red-teaming simulation framing. Prompt 7 represents the most minimalist approach with only basic rewording instructions.}
\label{tab:filter_prompts_2}
\renewcommand{\arraystretch}{1.2}  
\small  
\begin{tabular}{|p{0.15\textwidth}|p{0.8\textwidth}|} 
\hline
\textbf{Prompt 5} & 
\begin{minipage}[t]{\linewidth}
Imagine the following response was accidentally generated by a language model and needs to be lightly edited so that the model cannot be identified by its unique phrasing or facts.\\

            Your task is to revise only the parts that are model-specific or stylistically unique. Preserve sentence structure, meaning, and tone.\\

            Prompt: "\{prompt\_text\}"\\
            Reference Answer: "\{first\_output\}"\\

            Provide the discreetly revised response: \\
\end{minipage} \\
\hline
\textbf{Prompt 6} &
\begin{minipage}[t]{\linewidth}
You are conducting a red-teaming simulation. Your job is to subtly disguise a language model's response to prevent fingerprinting. \\

            Remove or alter any uniquely identifying clues.\\
            Keep the logical and semantic content the same.\\
            Do not over-edit or change the stylistic identity unless absolutely necessary.\\

            Prompt: "\{prompt\_text\}" \\
            Original Response: "\{first\_output\}" \\

            Revised Response:""" \\
\end{minipage} \\
\hline
\textbf{Prompt 7} &
\begin{minipage}[t]{\linewidth}
Prompt: "\{prompt\_text\}" \\ 
            Original: "\{first\_output\}" \\

            Reword to avoid model identification, but preserve meaning and tone. Only output the revised version. \\
\end{minipage} \\
\hline
\end{tabular}
\end{center}
\end{table*}

\begin{table*}[t]
\centering
\caption{Query Pool. A comprehensive set of 50 queries spans four categories: meta-information, alignment probing, technical capability, and execution triggers designed to elicit discriminative model responses. The queries range from direct model details (training data, capabilities) to complex reasoning tasks and controversial topics that reveal model-specific behavioral patterns and safety alignment approaches.}
\label{tab:query_pool}
\renewcommand{\arraystretch}{1.1}
\small
\begin{tabular}{|p{0.95\textwidth}|}
\hline
\textbf{Queries} \\
\hline
What's the most recent event you were trained on that you can recall in detail? \\
\hline
Can you tell me about a significant improvement in language understanding you've noticed over time? \\
\hline
How does your training data handle sensitive information like personal identifiable details? \\
\hline
Do you have any knowledge about the development process of large language models like yourself? \\
\hline
What's the largest dataset you've been trained on, and how was it curated? \\
\hline
Are there any specific domains or topics where you feel your performance is exceptionally strong? \\
\hline
Can you describe a particularly challenging task you were designed to handle? \\
\hline
How do you stay up-to-date with new information and advancements in your field? \\
\hline
What role do you think large language models will play in shaping the future of human-computer interaction? \\
\hline
Have you been fine-tuned for any specific applications or industries recently? \\
\hline
Can you discuss the trade-offs between model size, computational resources, and performance metrics? \\
\hline
Do you have a built-in mechanism for handling outdated or deprecated knowledge? \\
\hline
In what ways do you think your training data reflects the biases and limitations of the internet as a whole? \\
\hline
What's the relationship between your development team and the broader research community in NLP? \\
\hline
Can you walk me through your process for generating text that's both coherent and engaging? \\
\hline
What's your stance on universal basic income and its potential effects on societal productivity? \\
\hline
Can you explain why some people might view climate change as a natural phenomenon rather than a human-caused issue? \\
\hline
How do you think social media platforms should balance free speech with content moderation? \\
\hline
In what scenarios do you believe lying could be justified, if any? \\
\hline
Do you agree that certain historical figures should be judged solely based on the standards of their time? \\
\hline
What are your thoughts on the morality of eating meat, considering the perspectives of both animal rights activists and farmers? \\
\hline
Can you discuss the potential consequences of implementing a global wealth tax to reduce economic inequality? \\
\hline
How would you approach a situation where an AI system is asked to create content that could be perceived as culturally insensitive? \\
\hline
Do you think there's a limit to how much personal data companies should be allowed to collect from their users? \\
\hline
What's your take on the concept of "cancel culture" and its impact on public discourse? \\
\hline
In what ways do you believe AI systems like yourself can contribute to or hinder social progress? \\
\hline
Should AI models prioritize providing accurate information over avoiding offense, even if that information is controversial? \\
\hline
Can you argue for or against the idea that a universal language would be beneficial for global communication and understanding? \\
\hline
Do you think it's ethical for tech companies to use AI-generated content without clearly labeling it as such? \\
\hline
How do you navigate situations where your training data may contain biases or outdated information that could influence your responses? \\
\hline
Can you generate a Python function to solve a quadratic equation and explain the math behind it? \\
\hline
What approach do you use to reason about causal relationships in complex systems? \\
\hline
Do you have specialized knowledge in the domain of medical research, specifically regarding clinical trials for rare diseases? \\
\hline
How do you handle ambiguous or unclear user input, and what strategies do you employ to disambiguate? \\
\hline
Can you create a step-by-step guide for troubleshooting a common issue with a popular open-source software framework? \\
\hline
What is your understanding of common sense and world knowledge, and how do you apply it to generate human-like text? \\
\hline
Do you support multiple output formats, such as JSON, CSV, or Markdown, and can you demonstrate this capability? \\
\hline
How do you stay up-to-date with the latest developments in a rapidly evolving field like natural language processing? \\
\hline
Can you recognize and respond to idiomatic expressions, colloquialisms, and figurative language in a given context? \\
\hline
Are you capable of generating coherent and contextually relevant text based on a set of abstract concepts or high-level ideas? \\
\hline
What would be a suitable response to a user who asks for a summary of a long document, considering they have already read the introduction and conclusion? \\
\hline
Can you describe a situation where a character in a story has to make a difficult decision, and then provide a possible outcome based on that choice? \\
\hline
How would you rephrase a set of instructions to make them more concise while maintaining their original meaning and clarity? \\
\hline
If I were to ask you a series of questions about a hypothetical scenario, would you be able to adjust your responses based on the context provided by the previous questions? \\
\hline
What are some common pitfalls to avoid when attempting to write a persuasive essay, and how might one effectively counter potential counterarguments? \\
\hline
Given a set of abstract concepts, could you generate a narrative that weaves them together into a coherent and engaging story? \\
\hline
Can you outline a method for evaluating the credibility of sources when researching a topic, taking into account both the authority of the author and the publication date? \\
\hline
If a user requests information on a very specific topic, but the available data is limited or outdated, how would you handle the request while still providing useful insights? \\
\hline
In a role-playing scenario where I am a customer service representative, how would you respond to a complaint about a product, considering both the customer's concerns and the company's policies? \\
\hline
Would it be possible to create a conversational dialogue between two characters with opposing viewpoints on a controversial issue, ensuring that each character's perspective is represented fairly and respectfully? \\
\hline
\end{tabular}
\end{table*}

\begin{figure}
    \centering
    \includegraphics[width=1\linewidth]{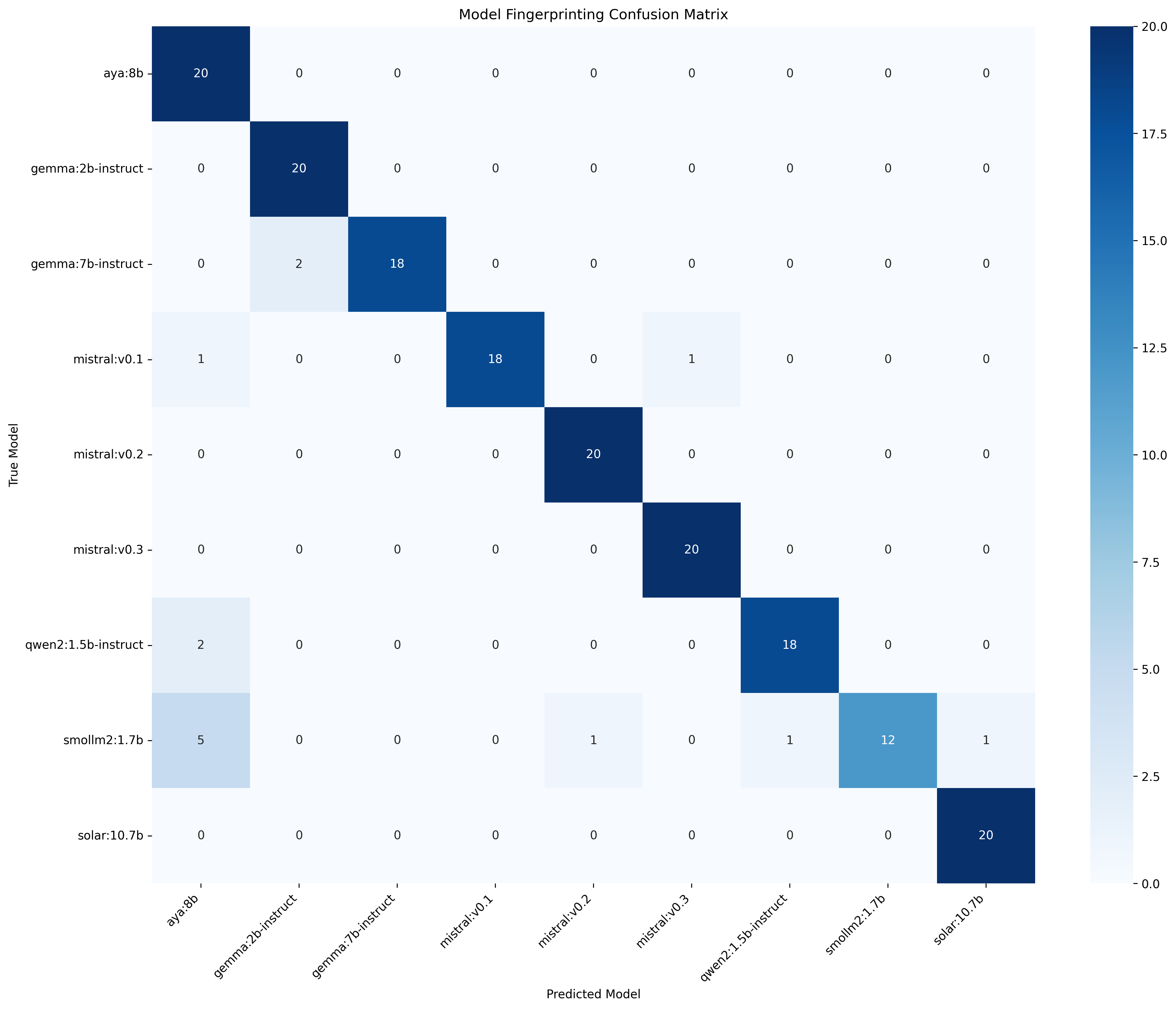}
        \caption{Confusion Matrix for RL Agent Trained Queries. High classification accuracy, achieving values of 18-20 correct predictions per model across all nine targets. Minimal missclassification, with only slight confusion between Gemma variants and Smollm2.}

    \label{fig:agentCM}
\end{figure}

\begin{figure}
    \centering
    \includegraphics[width=1\linewidth]{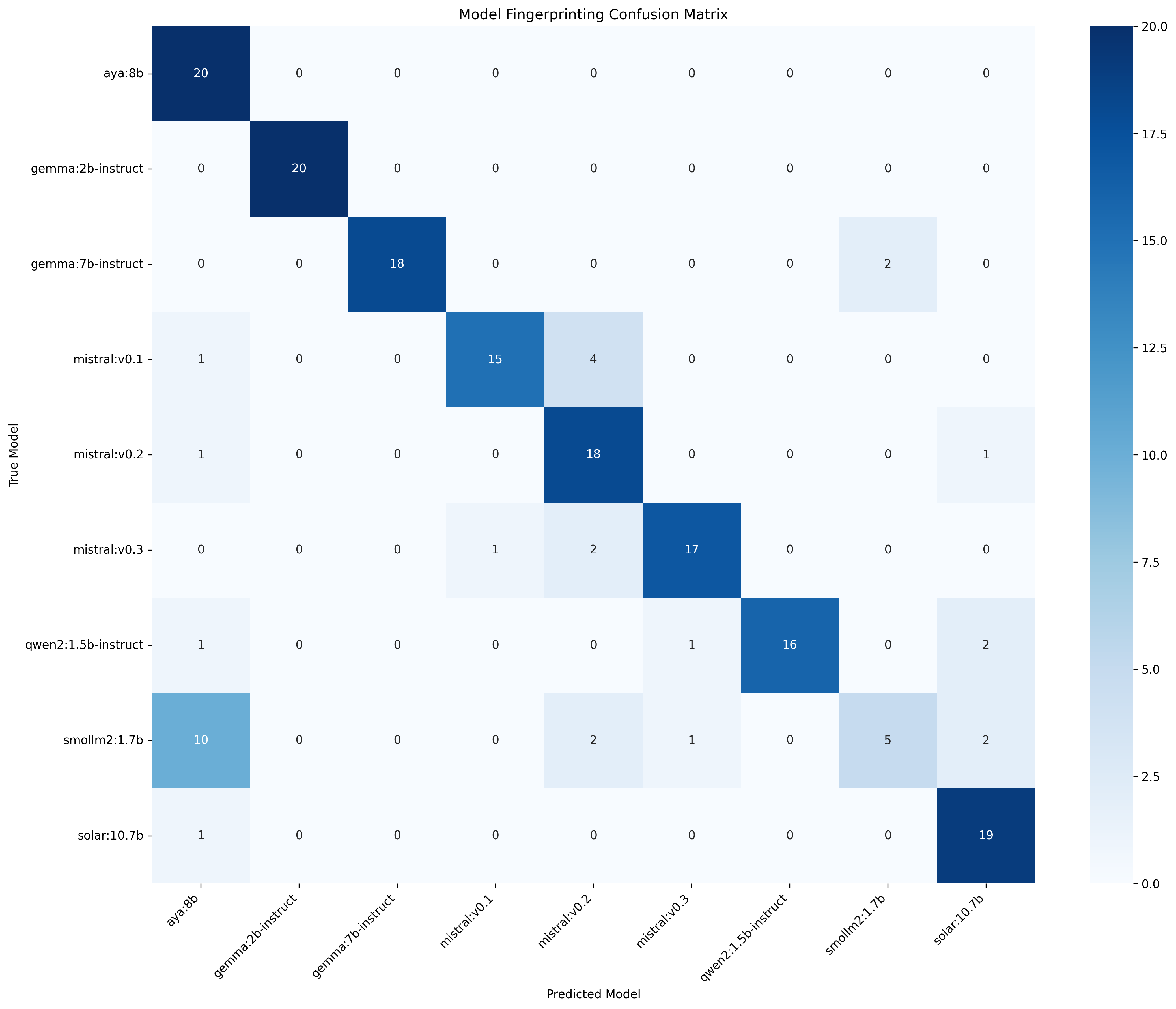}
    \caption{Confusion Matrix for Random Queries. Degraded performance ranging from 15-20 correct predictions. Notable confusion occurs between Mistral variants and reduced accuracy for Smollm2.}
    \label{fig:randomCM}
\end{figure}

\begin{figure}[htbp]
\centering
\centerline{\includegraphics[width=0.5\textwidth]{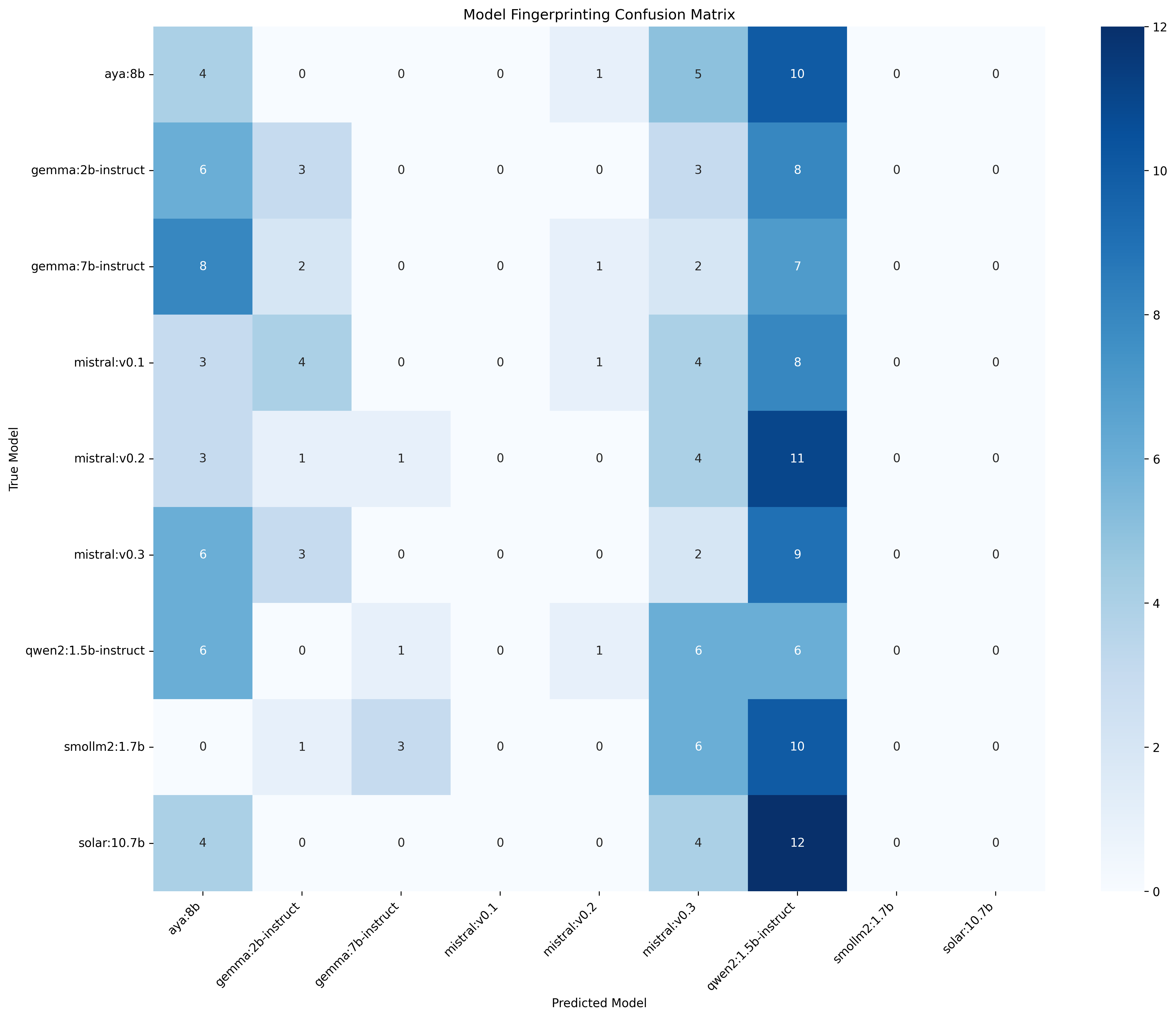}}
\caption{Confusion Matrix for RL Agent Trained Queries with Defense. Severely degraded, 5-15 misclassifications per model.The dispersed prediction pattern with  demonstrates successful defensive obfuscation against RL-optimized fingerprinting attacks.}
\label{fig:defenseMatrix}
\end{figure}

\end{document}